\documentclass[%
 reprint,
superscriptaddress,
amsmath,amssymb,
aps,
prl,
]{revtex4-1}

\usepackage{float}
\usepackage{graphicx}
\usepackage{dcolumn}
\usepackage{bm}

\usepackage{hyperref}
\hypersetup{
    colorlinks = true,
    urlcolor   = blue,
    citecolor  = red,
}

\begin{document}


\title{Helical Locomotion in Yield Stress Fluids}

\author{Farshad Nazari}
\affiliation{Department of Chemical and Biomedical Engineering, FAMU-FSU College of Engineering, Tallahassee, FL, 32310, USA}
\author{Kourosh Shoele}
\affiliation{Department of Mechanical Engineering, FAMU-FSU College of Engineering, Tallahassee, FL, 32310, USA}
\author{Hadi Mohammadigoushki}
\email[Corresponding Author: ]{hadi.moham@eng.famu.fsu.edu}
\affiliation{Department of Chemical and Biomedical Engineering, FAMU-FSU College of Engineering, Tallahassee, FL, 32310, USA}

\date{\today}

\begin{abstract}
\noindent We report three stages for locomotion of a helical swimmer in yield stress fluids. In the first stage, the swimmer must overcome material's yield strain to generate rotational motion. However, exceeding the first threshold is not sufficient for locomotion. Only when the viscous forces are sufficiently strong to plastically deform the material to a finite distance away from the swimmer, net locomotion will occur. Once locomotion is underway in the third stage, the yield stress retards swimming at small pitch angles. Conversely, at large pitch angles, yield stress dominates the flow by enhancing swimming speed. Flow visualizations reveal a highly localized flow near the swimmer in yield stress fluids.

 \end{abstract}

\maketitle

{ Microorganism’s locomotion is important in our daily life, environment, and physiology~\cite{Lau09}. For example, penetration of Helicobacter (H)-pylori through gastric mucus may cause ulcers and cancer~\cite{Miche02,Cel07,Ban13}, burrowing nematode worms through wet soil can enhance soil aeration and fertility~\cite{Fer96,Ghe01}, motility of bacteria may infect food products, and hydrogels that are used as tissue scaffolds, biofilms~\cite{FISCHER2015198,Kan18} or gives rise to a new set of advanced biosensors and biofilters~\cite{Vhu22,Gil21}. Rheological measurements indicate that these fluidic environments display a strong yield stress behavior~\cite{Cel07,Ghe01,Mouser2016}. Therefore, motility of organims in yield stress fluids is important in a host of applications, and its mechanistic understanding provides fundamental insights that can inform scientists on how to mitigate those health risks or design and engineer new materials for advanced applications.} Despite admirable progress in our understanding of locomotion in polymeric fluids~\cite{Li2021,PAT16,Spa13,Tho14,Liu11,She11,Lau07,Ter10,Bin19,Patt15}, little is known about locomotion in yield stress fluids. \par 

Yield stress materials behave like a solid and barely deform below the yield stress threshold. A recent experimental study showed that while at high pH (near neutral), H-pylori swims in porcine gastric mucus (PGM) freely~\cite{Ban13,Celli09}, at low pH (at which PGM is a yield stress fluid~\cite{Cel07,Xin99}), H-pylori is stuck in PGM~\cite{Ban13,Celli09}. A relevant theoretical study of Balmforth and co-workers showed that below a critical Bingham number, the yield stress impedes the locomotion of a 2D undulatory swimmer near a solid boundary in a simple Bingham fluid model~\cite{Bal10}. The Bingham number is defined as $Bi = {\sigma_y}/{\eta \dot{\gamma}} $, where $\sigma_y$ and $\eta$ and $\dot{\gamma}$ denote the yield stress, viscosity and the rate of deformation. Although these studies hint at existence of some critical thresholds that must be overcome by the organisms to gain motility in yield stress materials, such critical thresholds have not been quantified in experiments. Furthermore, only a limited theoretical effort has been devoted to probing swimming mechanisms, post yielding, in yield stress fluids~\cite{Hew17,Hew18,hewitt_balmforth_2022,eastham2022squirmer}. Particularly, Hewitt $\&$ Balmforth developed a slender body theory for yield stress fluids, and found that the optimal pitch angle associated with maximum swimming speed is moderately larger than the calculated one for the Newtonian fluids~\cite{Hew18,hewitt_balmforth_2022}. More recently, Eastham et al. investigated locomotion of a squirmer in a Bingham fluid model~\cite{eastham2022squirmer}. Despite these advances, there is a dearth of experimental studies that address the mechanisms of locomotion in yield stress fluids. Here we present the first experimental investigation of locomotion in yield stress fluids. We identify and document, for the very first time, two critical thresholds that must be overcome by the swimmer in order to propel itself forward in yield stress fluids. Furthermore, we show that the yield stress enriches the fluid dynamics once swimming is underway with a complex and localized yielded zone near the swimmer surface accompanied by an un-yielded region at a finite distance away from the swimmer.
 
Inspired by microswimmers such as H-pylori, we perform experiments with a 3D printed helical (corkscrew) swimmer. All swimmer's dimensions are held fixed except for the tail pitch angle which varies as $\psi = 12-74 ^{\circ}$. The swimmer is actuated via a uniform magnetic field of a rotating Holmholtz coil (This setup is similar to the one used in our previous study~\cite{Wu2021b} and shown in Fig. S1 of the supplementary information; SI). A combination of particle tracking velocimetry (PTV) and particle image velocimetry (PIV) is utilized to measure the swimming speed and to visualize the flow field around the swimmer. Yield stress fluids based on Carbopol solutions are considered. Additionally, Newtonian fluids based on corn-syrup are prepared for the purpose of comparison with the results obtained in yield stress fluids. The rheological properties of these fluids are reported in Table (S1) and Fig.~S2 of the SI. \par 

\begin{figure}[]
\centering

  \includegraphics[width=0.45\textwidth]{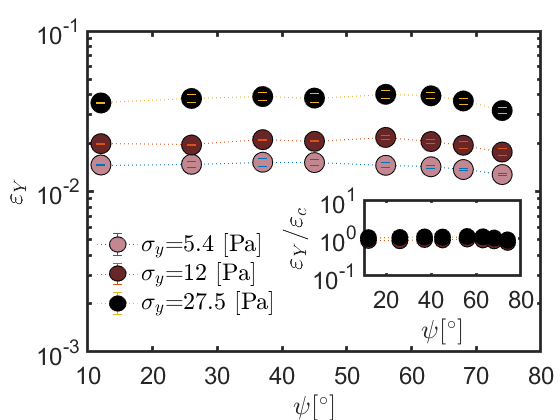}
   \caption{\small The critical yield strain as a function of swimmer's pitch angle in various yield stress fluids. The inset shows the ratio of yield strain obtained from swimming experiments to those measured in a commercial rheometer.}
    \label{strain}
\end{figure}

\begin{figure}[H]
\centering
\includegraphics[width=0.4\textwidth]{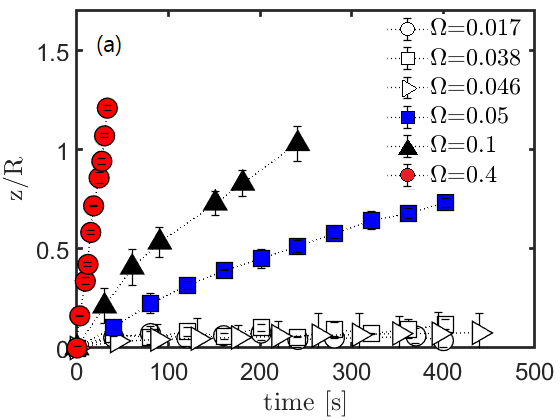}
	\includegraphics[width=0.4\textwidth]{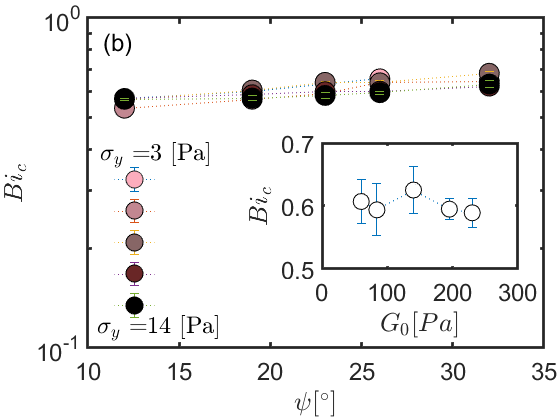}
\caption{\small (a) The trajectory of a swimmer with a pitch angle $\psi = 26 ^{\circ}$ as a function of time for different imposed rotation rates in a fluid with $\sigma_y = 5.4$ [Pa]. (b) The critical Bingham number associated with the onset of swimming as a function of pitch angle $\psi[^{\circ}]$
for various yield stress fluids. The inset shows the critical Bingham number as function of plateau modulus $G_0$.}
\label{Bic}

\end{figure}

The locomotion of the helical swimmer in yield stress fluids can be divided into three stages. The first stage is associated with a minimum torque needed to generate rotational motion. Below a critical torque ($\mathcal{T}_Y$), the swimmer does not have enough power to overcome the elastic resistance of the yield stress medium, and therefore, cannot rotate with the rotating magnetic field of the Helmholtz coil (see movie 1 and Fig.~S3 in the SI). Our hypothesis is that in stage (I), the yield stress around the swimmer transitions from a fully recoverable elastic network to a permanently plastic fluid and that triggers rotational motion. To test this hypothesis, we define a yield strain as: 
\begin{equation}
    \begin{aligned}
      \varepsilon_Y = \frac{\tau_Y}{G_o}, 
    \end{aligned}
     \vspace{-0.2cm}
    \end{equation}
\noindent where $\tau_Y$ and $G_o$ are the stress needed for the material to yield around the swimmer and the elastic modulus of the material. Here $\tau_Y \cong \frac{\mathcal{T}_Y}{A.L}$, where $A$ and $L$ are the total surface area and length of the swimmer (see caption of Fig.~S3 in the SI). Fig.~\ref{strain} shows that while the minimum yield strain needed to initiate rotational motion is independent of the swimmer's pitch angle, it increases by the yield stress of the material. To further analyze this data, we measured the yield strain of these fluids via a commercial rheometer ($\varepsilon_c$ see Fig.~S4 in SI). The inset of Fig.~\ref{strain} shows the ratio of these two yield strain values is remarkably close to unity. The latter result not only supports our hypothesis on the underlying cause of the first transition in swimming experiments, but also suggests that this swimmer may be used as an \textit{in situ} rheometer with potentially some broader impacts. For example, a magnetically actuated helical micro-swimmer robot could be used as an \textit{in situ} rheometer to characterize the yield strain of valuable biological gels and tissues that are not accessible in large quantity for bulk rheological measurements. 
Exceeding the first threshold is yet not sufficient for locomotion. In fact, for $\varepsilon > \varepsilon_Y$, and below a critical rotational velocity ($\Omega <\Omega_c$), the swimmer enters a second stage at which, despite in place rotational motion, net locomotion cannot be achieved (see Fig.~\ref{Bic}(a)). We checked that this critical rotational velocity is a strong function of yield stress and the swimmer's pitch angle (shown in Fig. S6 of the SI). Our hypothesis is that this critical threshold is controlled by a balance between viscous and yield stresses, which is captured through the Bingham number defined here based on the shear-thinning viscosity $\eta(\dot{\gamma})$, and a characteristic shear rate of $\dot{\gamma} = R\Omega/R = \Omega$, with $R$ being the radius of swimmer's cross-section. Fig.~\ref{Bic}(b) shows that through a Bingham number we can collapse all critical rotational velocities into a single graph over a broad range of swimmer pitch angles, and fluid rheological properties. Note that above this critical Bingham $Bi_c \approx 0.6$, a rotating swimmer does not generate net locomotion. Finally, swimmers with larger pitch angles ($\psi \ge 37 ^{\circ}$) always undergo locomotion even at the lowest accessible imposed rotation rates ($\Omega_{min}= 4\times 10^{-3}$ Hz). \par

\begin{figure}
\centering
	\includegraphics[width=0.43\textwidth]{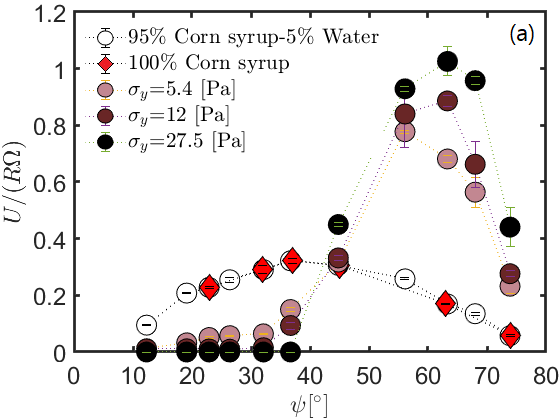}
 \includegraphics[width=0.44\textwidth]{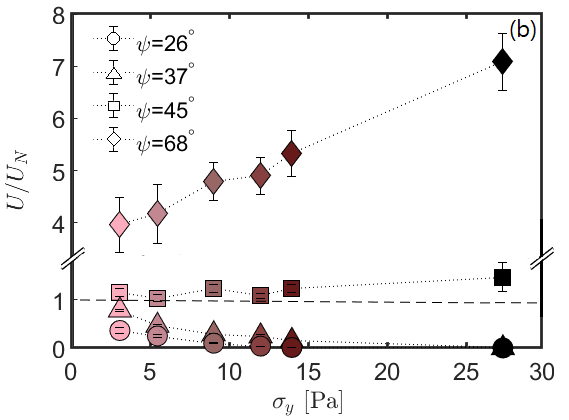}
	\caption{\small (a) Normalized swimming speed $U/(R{\Omega})$ as a function of pitch angle $\psi[^{\circ}]$ for various swimmers in yield stress fluids and the Newtonian fluids. (b) Swimming speed normalized by the Newtonian speed ($U/U_N$) as a function of $\sigma_y$ for swimmers with different pitch angles. In (b), the comparison is performed at the same imposed rotational velocity. }
	\label{Speed}
\end{figure}

By overcoming the above two critical thresholds ($\varepsilon_Y$ and $Bi_c$), the swimmer enters a stage (III) of locomotion, at which it is propelled at a constant velocity $U$. Fig.~\ref{Speed}(a) shows the normalized swimming speed ($U/R \Omega$) as a function of pitch angle for the Newtonian and sample yield stress fluids for $Bi< Bi_c$. Below the critical Bingham number, we observe that $U/R \Omega$ in the yield stress fluids is constant at different imposed rotation rates (see Fig. S6 in the SI). Therefore, Fig.~\ref{Speed}(a) shows the averaged swimming speed over a broad range of imposed rotation rates above $\Omega_c$. Additionally, we confirmed that $U/R \Omega$ is independent of the viscosity of the Newtonian fluid consistent with the resistive force theory~\cite{lauga_2020,Gray55}. Fig.~\ref{Speed}(a) provides three important and novel aspects of swimming in yield stress fluids. First, at low pitch angles ($ 12^{\circ} \le \psi \le 37 ^{\circ}$), the swimming speed in the yield stress fluids is lower than those measured in the Newtonian fluid. As the yield stress increases in this range of pitch angles, the swimming speed further decreases and approaches zero (see Fig.~\ref{Speed}(b)). These yield stress fluids are strongly shear-thinning, and shear-thinning has been known to enhance swimming speed of helical swimmers~\cite{Ard15,Gom17}. Hence, for these pitch angles, the yield stress dominates the flow by resisting against locomotion and shear-thinning effects are negligible. \par 

For $\psi = 45 ^{\circ}$, the swimming speed in the yield stress fluid is similar to those measured in the Newtonian fluid and by increasing the yield stress this ratio changes modestly (see Fig.~\ref{Speed}(b)). For larger pitch angles $\psi > 45 ^{\circ}$, the swimming speed in the yield stress fluid is much larger than those measured in the Newtonian fluids and increasing the yield stress further enhances the swimming speed (see Fig.~\ref{Speed}(b)). In shear-thinning fluids with a shear-thinning index of $n=0.47-0.9$, the maximum swimming enhancement has been noted to be about $50\%$~\cite{Gom17}. We surmise from these results that although shear-thinning might have contributed to swimming enhancement at large pitch angles, the swimming dynamics are primarily controlled by the yield stress. Finally, the optimal pitch angle in the yield stress fluid has shifted to much larger values compared to the Newtonian fluid (e.g., $\psi \approx 56^{\circ}$ and $\psi \approx 63^{\circ}$ for fluids with $\sigma_y$ = 5.4 and $\sigma_y$ = 27.5). The latter result is consistent with predictions of the slender body theory in the Bingham model~\cite{Hew18}. Note that Reynolds number is very small in these experiments and therefore, inertia does not play any role in these experiments (see Table~(S1) in the SI).\par

\begin{figure*}
    \centering
\includegraphics[width=0.95\textwidth]{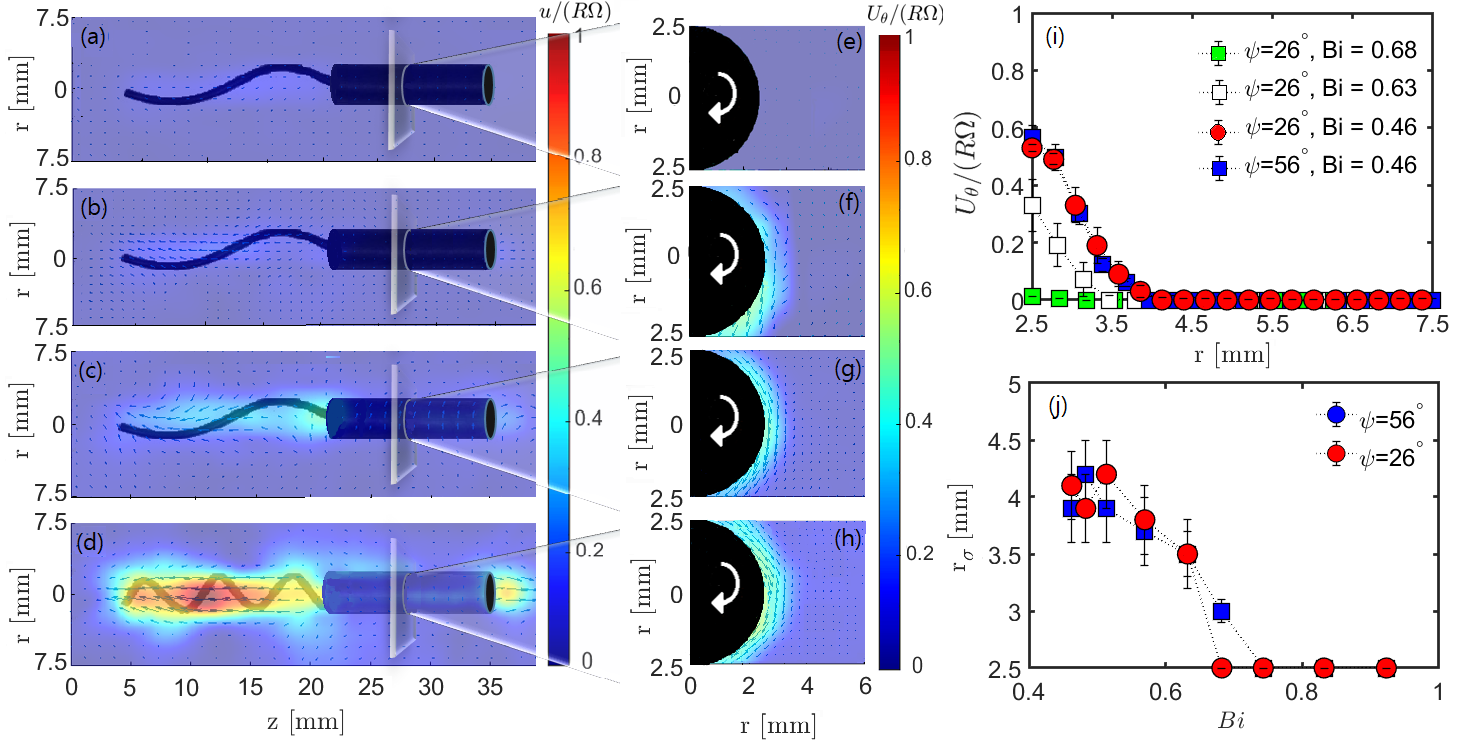}
	\caption{\small Time averaged 2D velocity profiles in the $r-z$ plane (a-d) and orthogonal to the direction of locomotion (e,h) in a yield stress fluid with $\sigma_y = 5.4$ [Pa]. In (e-h) the swimmer is rotating clock-wise. Panels (a,e), (b,f), (c,g) and (d,h) correspond to ($\psi = 26^{\circ}$, $Bi$ = 0.68), ($\psi = 26^{\circ}$, $Bi$ = 0.63), ($\psi = 26^{\circ}$, $Bi$ = 0.46) and ($\psi = 56^{\circ}$, $Bi$ = 0.46), respectively. (i) The time-averaged 1D angular velocity profile as a function of radial location around the head of the swimmer. (j) Location of the yield surface as a function of Bingham number for swimmers with different pitch angles. $u$ and $U_{\theta}$ denote the fluid velocity in the z and $\theta$ directions, respectively. Finally, the details of the time averaging are given in Fig.~S7 of the SI. }
\label{PIV}
\end{figure*}
To better understand the physics underlying the above stages of locomotion in yield stress fluids, we analyzed the detailed form of flow field around the swimmer. In stage (I), no net motion is observed. Hence, the fluid is stationary around the swimmer and PIV does not detect any motion. Fig.~\ref{PIV} (a-h) show a series of time-averaged two-dimensional velocity profiles around the swimmer in stage (II) and (III) of locomotion (see movie 2-3 in the SI). Recall that in stage (II), despite in place rotation, a swimmer can not generate net locomotion. In particular, Fig.~\ref{PIV}(a) shows that in stage (II) of locomotion and at high Bingham numbers ($Bi > Bi_c$; e.g., $Bi = 0.68$), while a weak propulsion is observed in the interior region of the helix in $r-z$ plane, the fluid around swimmer's head in the $r-\theta$ plane does not deform due to presence of a significant wall slip (albeit within the precision of our PIV measurements; see Fig.~\ref{PIV}(e,i)). This result is similar to experimental findings of Daneshi et al.~\cite{DANESHI201965} that showed capillary flows of yield stress fluids at low pressure gradients produce a fully plug flow with a significant wall-slip at the walls. As the Bingham number decreases towards $Bi_c$ (e.g., $Bi = 0.63$) and swimmer approaches stage (III) of its locomotion, the tail propulsion is boosted (shown in Fig.~\ref{PIV}(b)) and the yielded zone in $r-\theta$ plane is shifted to a finite distance away from swimmer's surface with a weaker wall slip (see Fig.~\ref{PIV}(f,i)). The corresponding flow fields in the Newtonian fluid are distinct from those measured in the yield stress fluids (see Fig.~S7 in the SI). While the velocity of the yield stress fluid decays quickly and approaches a non-yielded zone at a finite distance ($r_{\sigma}$) away from the swimmer (Fig.~\ref{PIV}(j)), the flow of a Newtonian fluid extends to much farther distances away from the swimmer, despite rotating at the same angular velocities. What these flow visualization findings reveal to us is that for $Bi > Bi_c$, the swimmer experiences an extremely confined space (and consequently a significant drag) around its head in yield stress fluids such that the weak fluid propulsion generated in the interior region of the helix is not strong enough to overcome this drag. Consequently, despite in place rotation, the swimmer does not progress forward in yield stress fluids. \par

Next, we investigate the impact of the tail pitch angle on the flow field around the swimmer and discuss the link between such flow fields and swimming speed results of Fig.~\ref{Speed}(a). For this purpose we consider two swimmers with $\psi = 26^{\circ}$ and $\psi = 56^{\circ}$ at $Bi = 0.46$ in stage (III) of their locomotion (cf., Fig.~\ref{PIV}(c) and Fig.~\ref{PIV}(d)). Fig.~\ref{PIV}(c,d) show that while the propulsion generated in the interior region of the helix is much stronger for the swimmer with a larger pitch angle, the velocity field in the $r-\theta$ plane (Fig.~\ref{PIV}(g,h,i)) and the location of the yield surface (Fig.~\ref{PIV}(j)) are the same for the two swimmers. These observations suggest that although swimmers experience a similar confinement (or drag) around their head, the swimmer with $\psi = 56^{\circ}$ experiences a stronger thrust and therefore, should be propelled faster than the swimmer with $\psi = 26^{\circ}$. This conclusion is consistent with the swimming speed data reported in Fig.~\ref{Speed}(a). Again, the corresponding experiments in the Newtonian fluid reveal a stark difference with those obtained in the yield stress fluid. The Newtonian fluid deforms to much farther distances away from the swimmer surface with a negligible wall-slip (see Fig.~S8 in the SI). These results highlight the strong coupling between a highly localized flow around the swimmer and the swimming dynamics in yield stress fluids.

In summary, we provided the first experimental investigation of the helical locomotion in yield stress fluids and illustrated that swimming can be divided into three stages. In the first stage, the swimmer must create rotational motion by overcoming the yield strain of the material ($\varepsilon_Y$). However, exceeding the first threshold is not sufficient for locomotion. For $Bi > Bi_c$, the yield stress fluid is hardly deformed around the swimmer, and the tail propulsion is not strong enough to generate locomotion. Only below a critical $Bi_c\approx 0.6$, when the rotational motion forces the material to yield at a finite distance away from the swimmer, forward motion will occur. Once swimming is underway in the third stage of locomotion, the swimming speed in the yield stress fluids is lower than those measured in Newtonian fluids for $\psi \leq 37^{\circ}$. Conversely, for $\psi \geq 45^{\circ}$ the swimmer moves faster in the yield stress fluids compared to their Newtonian counterparts. These observations open a new field of study and to fully understand them, there are still several questions that need to be answered, including why has the optimal pitch angle shifted to larger values in the yield stress fluid? What controls optimal swimming in yield stress fluids? What is the role of swimmer's head? How changing swimming strategy to a force/torque free mode affects the locomotion in yield stress fluids? We hope to address these questions in the near future.

We are grateful to Neil Balmforth and 
Philipe Coussot for several helpful discussions.


\end{document}